\documentclass[twoside,a4paper,11pt]{proceedings}
\usepackage{graphicx}
\usepackage{newtxtext,newtxmath}
\usepackage{hyperref}
\usepackage{movie15}
\usepackage{natbib}
\usepackage{float}
\newcommand{\HNII}{{\rm H}$\alpha+$[N {\sc ii}]~6548~\&~6584~\AA}

\begin{document}
\pagenumbering{arabic}
\pagestyle{myheadings}
\thispagestyle{empty}
\vspace*{-1cm}
{\flushleft\includegraphics[width=3cm,viewport=0 -30 200 -20]{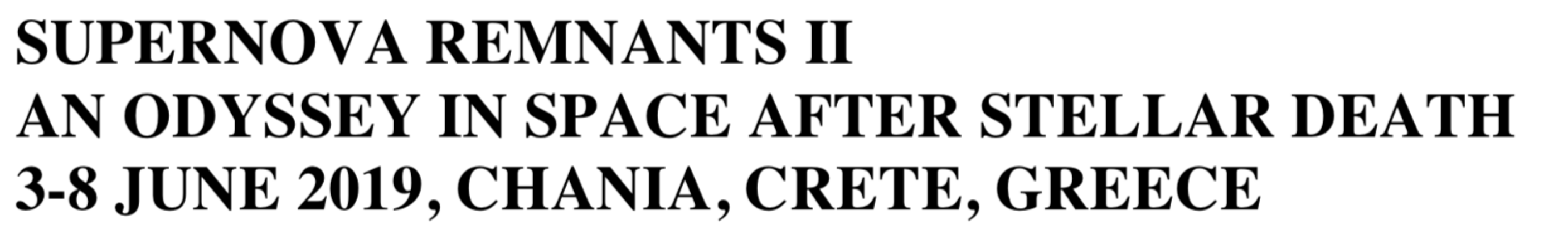}}
\vspace*{0.2cm}
\begin{flushleft}
{\bf {\LARGE
The first 3D Morpho-Kinematical model of a Supernova Remnant: The case of VRO 42.05.01 (G 166.0+4.3)}\\
\vspace*{1cm}
S. Derlopa$^{1,2}$
P. Boumis$^1$,
A. Chiotellis$^1$
W. Steffen$^3$ and
S. Akras$^4$
%
}\\
\vspace*{0.5cm}
%
$^{1}$
Institute for Astronomy, Astrophysics, Space Applications
and Remote Sensing, National Observatory of Athens,
15236 Penteli, Athens, Greece \\
$^{2}$
Department of Physics, University of Athens, Greece \\
$^{3}$
Instituto de Astronom\'ia, Universidad Nacional Aut\'onoma de M\'exico, Ensenada 22800, Baja California, Mexico \\
$^{4}$
Instituto de Matem\'atica, Estat\'istica e F\'isica, Universidade Federal do Rio Grande, Rio Grande 96203-900, Brazil
%
\end{flushleft}
\markboth{
 The first 3D morpho-kinematical model of a SNR: The case of VRO 42.05.01 (G 166.0+4.3)
}{
Derlopa et al.
}
\thispagestyle{empty}
\vspace*{0.4cm}
\begin{minipage}[l]{0.09\textwidth}
\ 
\end{minipage}
\begin{minipage}[r]{0.9\textwidth}
\vspace{1cm}
\section*{Abstract}{\small
We present the preliminary results of the first 3-dimensional (3D) Morpho-Kinematical (MK) model of a supernova remnant (SNR), using as a study case the Galactic SNR VRO 42.05.01 (G 166.0+4.3). For the purpose of our modelling, the astrophysical software SHAPE was employed in which wide field imaging and high resolution spectroscopic data were utilized. We found that the remnant is consisted by three basic distinctive components: a ``shell'', a ``wing'' and a ``hat'', which present different morphological and kinematical behaviour, probably due to different ambient medium properties. The whole nebula has an inclination of $6-8$ degrees with respect to the plane of the sky and a systemic velocity $V_{\rm sys}$ $\sim$ -15 to -25 \,km\,s$^{-1}$. Finally, we discuss the possible implications of our model's results on the origin and evolution of VRO 42.05.01.
 
\normalsize}
\end{minipage}

\section{Introduction}

Supernova Remnants (SNRs) are the diffuse nebulae that remain after the violent death of certain stars (progenitor stars). From the observational data of these objects we can trace back to the nature and evolution of the progenitor star. On the other hand, we can gain vital information about the properties of the ambient Circumstellar (CSM) and Interstellar (ISM) medium that the SNRs encounter during their evolution.

The fact that the observational images from the telescopes are 2D is a restrictive factor for the  thorough perception of the SNRs' properties. Important and useful information that could cover this gap can be provided through a 3D Morpho-Kinematical (MK) model, which is a 3D reconstruction of the morphology and kinematics of the studied nebula based on imaging and spectroscopic observational data. 

Up to date, all the 3D MK models that have been produced, regard Planetary Nebulae. Here we present the first 3D MK model for a SNR. The selected object is the Galactic SNR VRO 42.05.01 (Fig. \ref{fig:3d1}a). This SNR, which was selected due to its intriguing morphology, is consisted of two main parts: a hemisphere at the NE called the ``shell'' and a larger, bow-shaped shell at the SW, called the ``wing'' \citep{LAN1982}. For our MK model we used the code SHAPE \citep{STE2011} in which we applied new optical observational data (imaging and spectroscopic; \citealt{BOUMIS2016,BOUMIS_SNRII_2019}).


\section{Modelling}
\label{Modelling}

\begin{figure}[h!]
\begin{center}
 \includegraphics[scale=0.28]{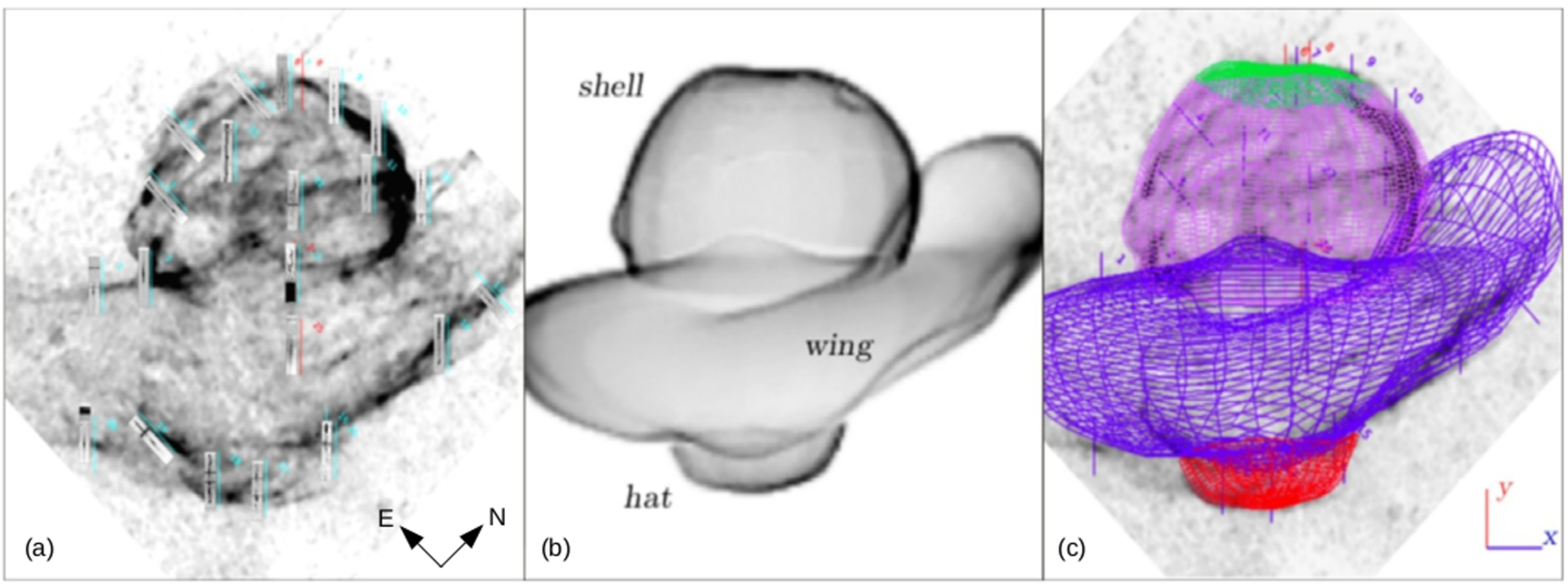}
 \caption{(a) \HNII~image of VRO 42.05.01 \citep{BOUMIS2016} along with the slits positions and the Position-Velocity diagrams in certain subregions of the nebula. (b) The 3D MK model of VRO 42.05.01 produced with the code SHAPE. c) Fig. \ref{fig:3d1}a image overlaid with the 3D model produced with SHAPE code in mesh-grid illustration. The colours correspond to distinctive components of VRO 42.05.01 in terms of kinematics.}
  \label{fig:3d1}
  \end{center}
  \end{figure}

\begin{figure}[h!]
\begin{center}
 \includegraphics[scale=0.15]{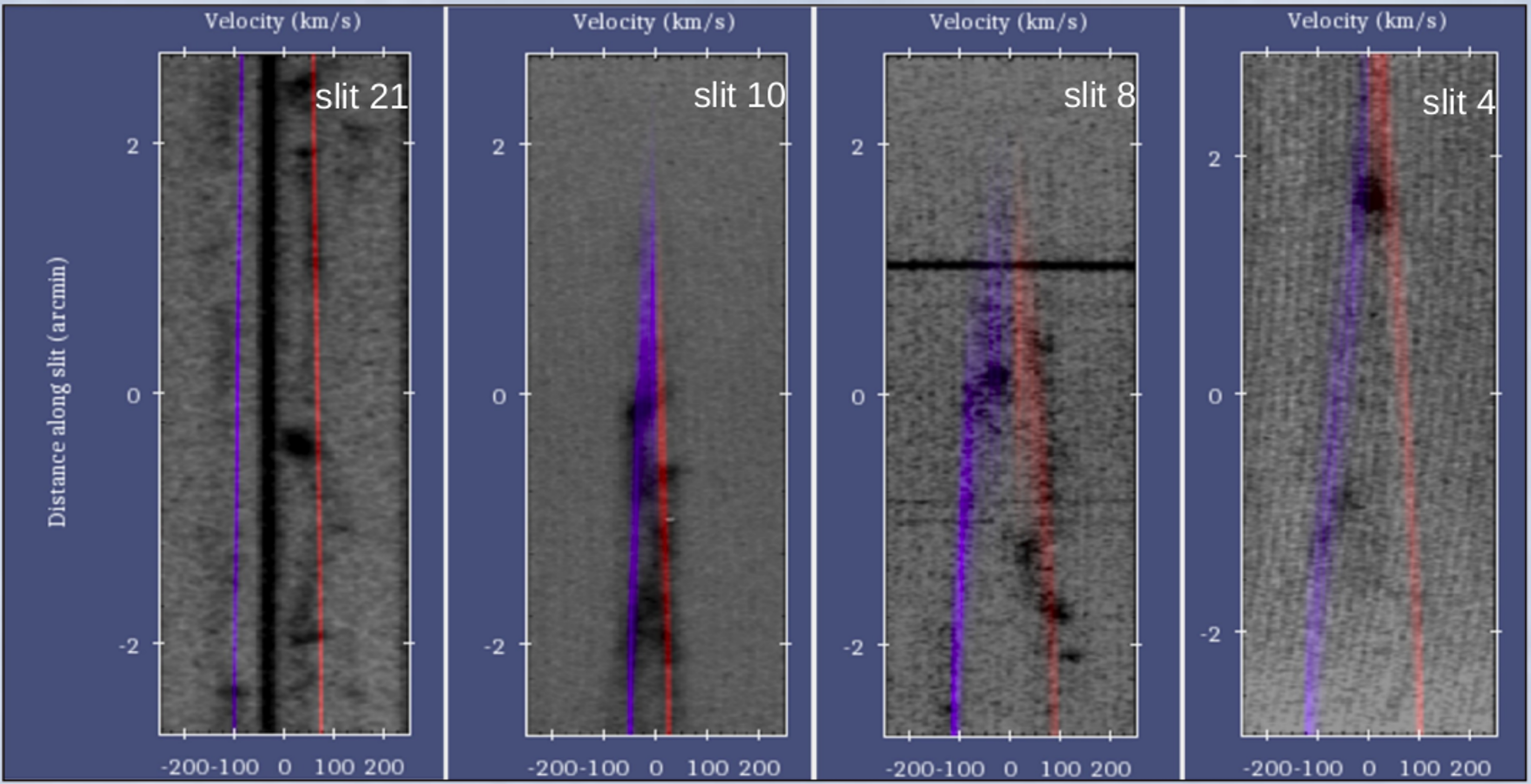}
 \caption{Observational PV diagrams (in black) from different areas of VRO, along with the synthetic PVs (blue/red shifted parts of the nebula) reproduced with the code SHAPE. The purpose is the best matching between observational and synthetic PVs, with respect to the overall structure of each spectrum.}
    \label{fig:slits}
    \end{center} 
  \end{figure}

For the 3D MK model of VRO 42.05.01 we employed the code SHAPE \citep{STE2011} in which we firstly applied wide field imaging data \citep{BOUMIS2016} of the remnant (insertion of morphological information) in order to reconstruct the 3D morphology of VRO 42.05.01, using the geometrical and physical modifiers provided by the code. Subsequently, the observational Position - Velocity (PV) diagrams deduced from the high resolution echelle spectra (Fig. \ref{fig:3d1}a) were imported in the code, providing the Doppler velocity for each subregion of the remnant (insertion of kinematical information). The next step was to create the synthetic PV diagrams - for the given SNR 3D morphology - using the code SHAPE and compare them with the observational PVs. In order to achieve the best agreement between the observational and synthetic PVs, we modified the initial 3D reconstruction of the object by altering the geometrical and physical modifiers, provoking by this way the change of the synthetic PVs. Fig. \ref{fig:3d1}b illustrates the 3D morphology of VRO 42.05.01 as resulted by our optimum model, while Fig. \ref{fig:slits} displays characteristic observational PV diagrams along with the synthetic PVs extracted by this model.

\section{Results}
\label{Results}
We found that VRO 42.05.01 is constituted by three main components, with regards to their morphology and kinematics: the ``shell'', the ``wing'' and the ``hat'' (Fig. \ref{fig:3d1}b). In particular we found the following properties for each component: i) the ``shell'' has an expansion velocity of $V_{\rm exp}$ $\sim$ 80-100~\,km\,s$^{-1}$. However, its upper part (Fig. \ref{fig:3d1}c, green region) deviates from the overall ``shell'''s expansion and expands with $V_{\rm exp}$ $\sim$ 180-200~\,km\,s$^{-1}$. This is the highest velocity that the remnant reveals. ii) The ``wing'', which is expanding slower than the ``shell'', reveals a velocity of $V_{\rm exp}$ $\sim$ 40-60~\,km\,s$^{-1}$. As far as its morphology is concerned, its right region has been tilted with respect to its left counterpart, while a part of the wing penetrates the ``shell'''s central region. iii) The third component of VRO, the ``hat'', which morphologically appears as a ``wing'''s protrusion, is also distinctive kinematically with a $V_{\rm exp}$ $\sim$ 80-100~\,km\,s$^{-1}$, which is different than the ``wing'''s $V_{\rm exp}$. Finally, we found that the whole remnant has a systemic velocity of $V_{\rm sys}$ $\sim$ -15 to -25 \,km\,s$^{-1}$, and an inclination of $6-8$ degrees, with the ``shell'' pointing inwards with respect to the plane of the sky.

\section{Discussion}
\label{Discussion}

We found that VRO 42.05.01 morphology comprises of three components: a ``shell'', a ``wing'' and a ``hat'', which apart from their clear morphological distinction, they also reveal different kinematical behaviour. In particular, the ``wing'' component, even if is way more extended that the ``shell'', it is expanding with a lower velocity than the latter. This fact points towards the conclusion that the ``wing'' was moving faster in a previous phase of the SNR evolution and at some point it has been substantially decelerated. In addition, the ``wing'' is tilted in its right portion, which indicates that this part encounters denser ambient medium. Another intriguing point of our results is that the ``hat'' and the upper part of the ``shell'' are expanding much faster than the components that they belong to. \citet{CHIOT_SNRII_2019} suggested that VRO 42.05.01 is shaped by the SNR interaction with an extended wind bubble, formed by the mass outflows of a runaway progenitor star. According to this model, the high velocities of the ``hat'' and the ``shell'''s upper part could be attributed to a shock breakout from the wind bubble to the surrounding ISM.

\begin{figure}
\center
\end{figure}

\small  
\section*{Acknowledgments}   
 S.D. and A.C. acknowledge the support of this work by the PROTEAS II project (MIS 5002515), which is implemented under the ``Reinforcement of the Research and Innovation Infrastructure" action, funded by the ``Competitiveness, Entrepreneurship and Innovation" operational programme (NSRF 2014-2020) and co-financed by Greece and the European Union (European Regional Development Fund). S.D acknowledges the Operational Programme ``Human Resources Development, Education and Lifelong Learning'' in the context of the project ``Strengthening Human Resources Research Potential via Doctorate Research'' (MIS-5000432), implemented by the State Scholarships Foundation (IKY) and co-financed by Greece and the European Union (European Social Fund- ESF). W.S. was supported by a grant from UNAM PAPIIT 104017. We would like to thank J. Dickel for his information about the origin of the name of VRO 42.05.01. VRO stands for the Vermilion River Observatory (University of Illinois), where the nebula was detected as a radio source at 610.5MHz in 1965 (\citealt{DICKEL1965}). This paper is based upon observations carried out at the Observatorio Astron\'{o}mico Nacional on the Sierra San Pedro M\'{a}rtir (OAN-SPM), BC, M\'{e}xico and Skinakas Observatory, Crete, Greece.

\bibliographystyle{aj}
\small
\bibliography{proceedings}

\begin{thebibliography}{}

\bibitem[\protect\citeauthoryear{{Boumis} et~al.}{{Boumis}
  et~al.}{2016}]{BOUMIS2016}
{Boumis}, P., {Akras}, S., {Leonidaki}, I., {Chiotellis}, A., {Kopsacheili},
  M., {Alikakos}, J., {Nanouris}, N.,  \& {Mavromatakis}, F. 2016, in Supernova
  Remnants: An Odyssey in Space after Stellar Death, 15

\bibitem[\protect\citeauthoryear{{Boumis} et~al.}{{Boumis}
  et~al.}{2019}]{BOUMIS_SNRII_2019}
{Boumis}, P., {Chiotellis}, A., {Derlopa}, S., {Akras}, S., {Leonidaki}, I.,
  {Alikakos}, J., {Kopsacheili}, M.,  \& {et al.} 2019, in Supernova Remnants
  II: An Odyssey in Space after Stellar Death, in press

\bibitem[\protect\citeauthoryear{{Chiotellis} et~al.}{{Chiotellis}
  et~al.}{2019}]{CHIOT_SNRII_2019}
{Chiotellis}, A., {Boumis}, P., {Derlopa}, S.,  \& {Steffen}, W. 2019, in
  Supernova Remnants II: An Odyssey in Space after Stellar Death, in press

\bibitem[\protect\citeauthoryear{{Dickel}, {McGuire}, \& {Yang}}{{Dickel}
  et~al.}{1965}]{DICKEL1965}
{Dickel}, J.~R., {McGuire}, J.~P.,  \& {Yang}, K.~S. 1965, apj, 142, 798

\bibitem[\protect\citeauthoryear{{Landecker} et~al.}{{Landecker}
  et~al.}{1982}]{LAN1982}
{Landecker}, T.~L., {Pineault}, S., {Routledge}, D.,  \& {Vaneldik}, J.~F.
  1982, apj, 261, L41

\bibitem[\protect\citeauthoryear{{Steffen} et~al.}{{Steffen}
  et~al.}{2011}]{STE2011}
{Steffen}, W., {Koning}, N., {Wenger}, S., {Morisset}, C.,  \& {Magnor}, M.
  2011, IEEE Transactions on Visualization and Computer Graphics, Volume 17,
  Issue 4, p.454-465, 17, 454

\end{thebibliography}

\end{document}